\documentclass[arps,prd,onecolumn,showpacs,preprintnumbers,letter`paper]{revtex4}

\usepackage{graphicx}
\usepackage{amssymb}
\usepackage{amsmath}
\usepackage{bm}
\usepackage{color}
\usepackage[utf8]{inputenc}
\usepackage{comment}

\definecolor{ultramarine}{rgb}{0.07, 0.04, 0.56}
\definecolor{cadmiumgreen}{rgb}{0.0, 0.42, 0.24}
\definecolor{indigo(dye)}{rgb}{0.0, 0.25, 0.42}
\usepackage[linktocpage=true]{hyperref}
\hypersetup{
colorlinks=true,
citecolor=ultramarine,
linkcolor=cadmiumgreen,
urlcolor=indigo(dye),
pdfauthor={},
pdftitle={},
pdfsubject={}
}

\def\be{\begin{equation}}
\def\ee{\end{equation}}
\def\ba#1\ea{\begin{align}#1\end{align}}
\newcommand{\vs}{\nonumber\\}

\newcommand{\refeq}[1]{Eq.~(\ref{eq:#1})}
\newcommand{\refeqs}[2]{Eqs.~(\ref{eq:#1})--(\ref{eq:#2})}
\newcommand{\refEq}[1]{Eq.~(\ref{eq:#1})}
\newcommand{\refEqs}[2]{Eqs~(\ref{eq:#1})--(\ref{eq:#2})}

\newcommand{\refsec}[1]{Sec.~\ref{sec:#1}}
\newcommand{\refapp}[1]{App.~\ref{app:#1}}

\renewcommand{\v}[1]{\mathbf{#1}}
\renewcommand{\vr}{\v{r}}

\newcommand{\vk}{\v{k}}
\newcommand{\vq}{\v{q}}
\renewcommand{\d}{\delta}

\newcommand{\dirac}{\delta_D}
\newcommand{\tpc}{(2\pi)^3}

\renewcommand{\[}{\left[}
\renewcommand{\]}{\right]}
\renewcommand{\(}{\left(}
\renewcommand{\)}{\right)}
\newcommand{\fnl}{f_{\rm NL}}
\newcommand{\Phh}{P_{hh}}
\newcommand{\Pdd}{P_{\delta\delta}}
\newcommand{\Ppp}{P_{\phi\phi}}
\newcommand{\Pdp}{P_{\delta\phi}}
\newcommand{\Bhhh}{B_{hhh}}
\newcommand{\Bhhhsq}{B_{hhh}^{\rm sq}}

\begin{document}

\preprint{YITP-SB-17-02}

\title{The halo squeezed-limit bispectrum with primordial non-Gaussianity: a power spectrum response approach}

\author{Chi-Ting Chiang}

\affiliation{C.N. Yang Institute for Theoretical Physics, Department of Physics \& Astronomy,
Stony Brook University, Stony Brook, NY 11794}

\begin{abstract}
Modeling the nonlinearity of the halo bispectrum remains a major challenge
in modern cosmology, in particular for ongoing and upcoming large-scale
structure observations that are performed to study the inflationary physics.
The ``power spectrum response'' offers a solution for bispectrum in the
so-called squeezed limit, in which one wavenumber is much smaller than
the other two. As a first step, we demonstrate that the halo squeezed-limit
bispectrum computed from the second-order standard perturbation theory
agrees precisely with the responses of \emph{linear} halo power spectrum
to large-scale density and potential fluctuations. Since the halo power
spectrum responses to arbitrarily small scales can straightforwardly be
obtained by separate universe simulations, the response approach provides
a novel and powerful technique for modeling the nonlinear halo squeezed-limit
bispectrum.
\end{abstract}

\maketitle

\section{Introduction}
\label{sec:intro}
The squeezed-limit bispectrum that quantifies the correlation between two
small-scale modes and one large-scale mode captures the impact of the
large-scale density environment on the small-scale power spectrum. This
coupling between large- and small-scale modes is generated by nonlinear
gravitational evolution, and possibly by the inflationary physics that
produces local non-Gaussianity in the primordial curvature perturbation
(see Ref.~\cite{Alvarez:2014vva} for a recent review). Measurement of
the squeezed-limit bispectrum can thus be used to test our understanding
of gravity and the physics of inflation.

Traditionally the bispectrum is computed with the perturbation theory
(see Ref.~\cite{Bernardeau:2001qr} for a review on standard perturbation
theory (hereafter SPT) and Ref.~\cite{Porto:2016pyg} for a review on
effective field theory), in which the matter or halo density perturbations
are expanded in series of the linear Gaussian perturbation, and the $n$-point
function can be calculated with the Wick's theorem. A novel and powerful
approach to computing the squeezed-limit bispectrum is to consider how the
small-scale power spectrum \emph{responds to} the large-scale environment
\cite{Li:2014sga,Chiang:2014oga}. We shall refer it as the ``response''
approach.

In this paper, we demonstrate that the halo squeezed-limit bispectrum
with local primordial non-Gaussianity can be derived from the responses
of halo power spectrum to the large-scale density and potential perturbations.
Specifically, we take the SPT framework and consider that halo number
density traces the underlying density and potential fluctuations, ignoring
the large-scale tidal field (see Ref.~\cite{Desjacques:2016bnm} for a
recent review). For simplicity, we assume that the large-scale halo biases
are \emph{local in Eulerian space}, but our derivation can be generalized
to local Lagrangian bias model with the transformation from Lagrangian to
Eulerian space (see e.g. Refs.~\cite{Giannantonio:2009ak,Baldauf:2010vn}).
Ref.~\cite{Baldauf:2015vio} has done a related work to measure how the halo
power spectrum responds to a long-wavelength matter fluctuation in $N$-body
simulations, but we shall make a clearer connection between the halo
squeezed-limit bispectrum and the halo power spectrum response as well
as extend to the cosmology with local primordial non-Gaussianity.

The rest of the paper is organized as follows.
In \refsec{bisp} we use the SPT to expand the halo number density fluctuation
in series of the underlying density and potential perturbations to the second
order, and show the leading-order halo squeezed-limit bispectrum.
In \refsec{resp} we compute the responses of the linear halo power spectrum
to large-scale density and potential perturbations and connect this result
to the halo squeezed-limit bispectrum.
We discuss the result and future applications in \refsec{discu}.
In \refapp{bhhh} we demonstrate the detailed derivation of the halo
squeezed-limit bispectrum using the second-order SPT.

\section{Second-order standard perturbation theory}
\label{sec:bisp}
In the peak-background split picture, the long-wavelength perturbations
change the local overdensity threshold for halo formation \cite{Slosar:2008hx}.
Therefore, in the large-scale limit, the halo number density is modulated
by the long-wavelength perturbations and can be expanded to the second
order as \cite{McDonald:2008sc,Giannantonio:2009ak,Baldauf:2010vn}
\be
 n_h(\vr|\d_l,\phi_l)=\bar{n}_h(\vr)
 +\frac{\partial\bar{n}_h}{\partial\d_l}\d_l(\vr)
 +\frac{\partial\bar{n}_h}{\partial\phi_l}\phi_l(\vr)
 +\frac{2}{2!}\frac{\partial^2\bar{n}_h}{\partial\d_l\partial\phi_l}\d_l(\vr)\phi_l(\vr)
 +\frac{1}{2!}\frac{\partial^2\bar{n}_h}{\partial\d_l^2}\d_l^2(\vr)
 +\frac{1}{2!}\frac{\partial^2\bar{n}_h}{\partial\phi_l^2}\phi_l^2(\vr) \,,
\label{eq:n_h}
\ee
where $\bar{n}_h$ is the mean halo number density, and $\d_l$ and $\phi_l$
are linear Gaussian density and potential fluctuations, respectively.
For simplicity, in \refeq{n_h} we neglect $\langle\d_l^2(\vr)\rangle$,
$\langle\d_l(\vr)\phi_l(\vr)\rangle$, and $\langle\phi_l^2(\vr)\rangle$,
which assure $\langle n_h(\vr|\d_l,\phi_l)\rangle=\bar{n}_h(\vr)$, as
they only contribute to the $k=0$ mode in Fourier space. We consider
that $\phi_l$ is the primordial potential in the matter-dominated epoch,
hence it is related to $\d_l$ by $\d_l(\vk,a)=M(k,a)\phi_l(\vk)$, with
the Poisson operator $M(k,a)=\frac23\frac{D(a)}{H_0^2\Omega_m}k^2T(k)$.
Here $a$ is the scale factor, $D$ is the linear growth normalized to $a$
in the matter-dominated epoch, $H_0$ is the present-day Hubble, $\Omega_m$
is the present-day fractional energy density of matter, and $T(k)$ is the
matter transfer function \footnote{Strictly speaking, $\phi_l$ is the Bardeen's
curvature perturbation instead of Newtonian potential due to the positive
sign in the Poisson operator. In this paper we loosely call $\phi_l$ as potential
perturbation.}. In the following, we shall omit the scale factor argument
for simplicity. Defining the halo bias as
\be
 b_{ij}=\frac{1}{\bar{n}_h}\frac{\partial^{i+j}\bar{n}_h}{\partial\d_l^i\partial\phi_l^j} \,,
\label{eq:bias}
\ee
we can write the halo number density fluctuation as
\be
 \d_h(\vr)=\frac{n_h(\vr|\d_l,\phi_l)}{\bar{n}_h}-1
 =b_{10}\d_l(\vr)+b_{01}\phi_l(\vr)+b_{11}\d_l(\vr)\phi_l(\vr)
 +\frac{b_{20}}{2}\d_l^2(\vr)+\frac{b_{02}}{2}\phi_l^2(\vr) \,.
\label{eq:d_h_1}
\ee
\refEq{bias} indicates that the biases are the responses of the halo mass
function to the large-scale density or potential fluctuations. Note that
for simplicity we consider both $\d_l$ and $\phi_l$ are in Eulerian space,
hence the biases are the Eulerian biases. Our derivation can be generalized
to the Lagrangian bias model by expanding $\d_h$ in Lagrangian space as
\refeq{d_h_1} and transforming $\d_h$ to Eulerian space using the conservation
of the number of halos (see e.g. Refs.~\cite{Giannantonio:2009ak,Baldauf:2010vn}).

In general, $\d_h$ traces the underlying \emph{nonlinear} matter density
fluctuation $\d_m$ and potential fluctuation $\Phi$, namely
\be
 \d_h(\vr)=b_{10}\d_m(\vr)+b_{01}\Phi(\vr)+b_{11}\d_m(\vr)\Phi(\vr)
 +\frac{b_{20}}{2}\d_m^2(\vr)+\frac{b_{02}}{2}\Phi^2(\vr) \,.
\label{eq:d_h_2}
\ee
In the presence of the local primordial non-Gaussianity, the potential
perturbation is $\Phi(\vr)=\phi_l(\vr)+\fnl\phi_l^2(\vr)$ \cite{Komatsu:2001rj},
where $\fnl$ quantifies the amount of local non-Gaussianity, with $\fnl=0$
being Gaussian. Using the Poisson operator, we can compute the matter density
up to the second order in Fourier space as
\be
 \d_m(\vk)=\d_l(\vk)+\fnl M(k)\int\frac{d^3q}{\tpc}~\phi_l(\vq)\phi_l(\vk-\vq)
 +\int\frac{d^3q}{\tpc}~\d_l(\vq)\d_l(\vk-\vq)F_2(\vq,\vk-\vq) \,,
\label{eq:d_m}
\ee
where the last term of \refeq{d_m} is due to the nonlinear gravitational
evolution with the kernel $F_2$ computed from SPT. Combining \refeqs{d_h_2}{d_m},
we have the halo number density fluctuation up to the second order in $\d_l$ and
$\phi_l$ as
\ba
 \d_h(\vk)\:&=b_{10}\[\d_l(\vk)+\int\frac{d^3q}{\tpc}~\d_l(\vq)\d_l(\vk-\vq)F_2(\vq,\vk-\vq)
 +\fnl M(k)\int\frac{d^3q}{\tpc}~\phi_l(\vq)\phi_l(\vk-\vq)\] \vs
 \:&+b_{01}\[\phi_l(\vk)+\fnl\int\frac{d^3q}{\tpc}~\phi_l(\vq)\phi_l(\vk-\vq)\]
 +b_{11}\int\frac{d^3q}{\tpc}~\d_l(\vq)\phi_l(\vk-\vq) \vs
 \:&+\frac{b_{20}}{2}\int\frac{d^3q}{\tpc}~\d_l(\vq)\d_l(\vk-\vq)
 +\frac{b_{02}}{2}\int\frac{d^3q}{\tpc}~\phi_l(\vq)\phi_l(\vk-\vq) \,.
\label{eq:d_h_3}
\ea

The halo power spectrum and bispectrum are defined as
\be
 \langle\d_h(\vk)\d_h(\vk')\rangle=\tpc\dirac(\vk+\vk')\Phh(k) \,, \quad
 \langle\d_h(\vk_1)\d_h(\vk_2)\d_h(\vk_3)\rangle=\tpc\dirac(\vk_1+\vk_2+\vk_3)\Bhhh(\vk_1,\vk_2,\vk_3) \,,
\label{eq:pkbk}
\ee
where $\dirac$ is the Dirac delta function. Inserting \refeq{d_h_3} into
\refeq{pkbk} as well as using the Wick's theorem, we obtain the leading
order halo power spectrum as (up to $\d^n\phi^m$ with $n+m=2$) 
\be
 \Phh(k)=b_{10}^2\Pdd(k)+b_{01}^2\Ppp(k)+2b_{10}b_{01}\Pdp(k) \,,
\label{eq:Phh}
\ee
where $\Pdd$, $\Ppp$, and $\Pdp$ are linear density-density, potential-potential,
and density-potential power spectra, respectively. For the halo squeezed-limit
bispectrum, we consider the configuration $k_1\approx k_2=k\gg k_3=k_L$. The
detailed derivation is given in \refapp{bhhh}, and at the leading order we have
(up to $\d^n\phi^m$ with $n+m=4$)
\be
 \Bhhhsq(\vk_1,\vk_2,\vk_3)=b_{10}f_\d(k)\Pdd(k_L)+b_{01}f_\phi(k)\Ppp(k_L)+\[b_{01}f_\d(k)+b_{10}f_\phi(k)\]\Pdp(k_L) \,,
\label{eq:Bhhh}
\ee
where
\ba
\label{eq:f_d}
 f_\d(k)\:&=2b_{10}b_{20}\Pdd(k)
 +b_{10}^2\[\frac{47}{21}-\frac13\frac{d\ln\Pdd}{d\ln k}\]\Pdd(k)
 +2b_{10}b_{11}\Pdp(k) \vs
 \:&+2b_{01}b_{20}\Pdp(k)+b_{10}b_{01}\[\frac{47}{21}-\frac13\frac{d\ln\Pdd}{d\ln k}\]\Pdp(k)
 +2b_{01}b_{11}\Ppp(k) \,, \\
 f_\phi(k)\:&=2b_{10}b_{11}\Pdd(k)+2b_{01}b_{11}\Pdp(k)
 +2b_{10}b_{02}\Pdp(k)+2b_{01}b_{02}\Ppp(k) \vs
 \:&+4\fnl b_{10}^2\Pdd(k)+4\fnl b_{01}^2\Ppp(k)+8\fnl b_{10}b_{01}\Pdp(k) \,.
\label{eq:f_phi}
\ea
\refEqs{Bhhh}{f_phi} are the primary results in this section, i.e. the halo
squeezed-limit bispectrum with primordial non-Gaussianity derived from the
second-order SPT. We shall show in \refsec{resp} that the same result can
be obtained by considering how \emph{linear} halo power spectrum responds
to large-scale density and potential fluctuations.

\section{Linear halo power spectrum response}
\label{sec:resp}
Let us now turn to the response approach. In the presence of $\d_l$ and $\phi_l$,
at the leading order the halo power spectrum is modulated as
\be
 \Phh(k|\d_l,\phi_l)=\left.\Phh(k)\right|_{\d_l,\phi_l=0}
 +\left.\frac{\partial\Phh(k)}{\partial\d_l}\right|_{\d_l,\phi_l=0}\d_l
 +\left.\frac{\partial\Phh(k)}{\partial\phi_l}\right|_{\d_l,\phi_l=0}\phi_l \,,
\label{eq:Phh_1}
\ee
where $\left.\Phh(k)\right|_{\substack{\d_l,\phi_l=0}}$ is given by \refeq{Phh},
and $\partial\Phh(k)/\partial\d_l$ and $\partial\Phh(k)/\partial\phi_l$ are the
responses of the halo power spectrum to large-scale density and potential perturbations,
respectively. Since the squeezed-limit bispectrum is essentially the coupling
between the small-scale power spectrum and its large-scale environment, we compute
the correlation between $\Phh(k|\d_l,\phi_l)$ and the large-scale halo density
fluctuation $(b_{10}\d_l+b_{01}\phi_l)$ (as the long-wavelength mode $k_3$ in
the squeezed-limit bispectrum calculation) and obtain
\be
 \langle\Phh(k|\d_l,\phi_l)(b_{10}\d_l+b_{01}\phi_l)\rangle=
 b_{10}\frac{\partial\Phh(k)}{\partial\d_l}\langle\d_l^2\rangle
 +b_{01}\frac{\partial\Phh(k)}{\partial\phi_l}\langle\phi_l^2\rangle
 +\[b_{01}\frac{\partial\Phh(k)}{\partial\d_l}+b_{10}\frac{\partial\Phh(k)}{\partial\phi_l}\]\langle\d_l\phi_l\rangle \,,
\label{eq:corr}
\ee
where $\langle\d_l^2\rangle=\Pdd(k_L)$, $\langle\phi_l^2\rangle=\Ppp(k_L)$,
and $\langle\d_l\phi\rangle=\Pdp(k_L)$ are the large-scale linear power
spectra. Note that \refeq{corr} has the same form as \refeq{Bhhh}.

There are three ways that the halo power spectrum responds to $\d_l$ and $\phi_l$.
The first way is through the halo bias, and it can be computed via \refeq{bias} as
\be
 \frac{\partial b_{10}}{\partial\d_l}=b_{20}-b_{10}^2 \,, \quad
 \frac{\partial b_{01}}{\partial\phi_l}=b_{02}-b_{01}^2 \,, \quad
 \frac{\partial b_{10}}{\partial\phi_l}=\frac{\partial b_{01}}{\partial\d_l}=b_{11}-b_{10}b_{01} \,.
\label{eq:b_resp}
\ee
Combining with \refeq{b_resp} with \refeq{Phh}, we have
\ba
 \left.\frac{\partial\Phh(k)}{\partial\d_l}\right|_{\substack{{\rm halo}\\{\rm bias}}}\:&=
 2(b_{10}b_{20}-b_{10}^3)\Pdd(k)+2(b_{20}b_{01}-2b_{10}^2b_{01}+b_{10}b_{11})\Pdp(k)
 +2(b_{01}b_{11}-b_{10}b_{01}^2)\Ppp(k) \,, \\
 \left.\frac{\partial\Phh(k)}{\partial\phi_l}\right|_{\substack{{\rm halo}\\{\rm bias}}}\:&=
 2(b_{10}b_{11}-b_{10}^2b_{01})\Pdd(k)+2(b_{01}b_{11}-2b_{10}b_{01}^2+b_{10}b_{02})\Pdp(k)
 +2(b_{01}b_{02}-b_{01}^3)\Ppp(k) \,.
\ea

The second way is through the small-scale linear power spectra. For the response
to $\d_l$, we consider that the effect is due to nonlinear gravitational evolution,
i.e. the squeezed-limit matter bispectrum in the absence of local primordial
non-Gaussianity, hence the potential power spectra do not respond to $\d_l$.
As a result, at the leading order we have (see e.g. Refs.~\cite{Li:2014sga,Chiang:2014oga})
\be
 \frac{\partial\Pdd(k)}{\partial\d_l}=\[\frac{47}{21}-\frac13\frac{d\ln\Pdd}{d\ln k}\]\Pdd(k) \,, \quad
 \frac{\partial\Pdp(k)}{\partial\d_l}=\frac12\[\frac{47}{21}-\frac13\frac{d\ln\Pdd}{d\ln k}\]\Pdp(k) \,, \quad
 \frac{\partial\Ppp(k)}{\partial\d_l}=0 \,.
\label{eq:pk_resp_d}
\ee
For the response to $\phi_l$, we consider that the effect is due to the initial
conditions, i.e. the primordial non-Gaussianity, hence all linear power spectra
respond identically. Following Ref.~\cite{Slosar:2008hx}, the effect of $\phi_l$
can be regarded as a change of $\sigma_8$ locally, and at the leading order we have
\be
 \frac{\partial P_{xy}(k)}{\partial\phi_l}
 =\frac{\partial P_{xy}(k)}{\partial\sigma_8}\frac{\partial\sigma_8}{\partial\phi_l}
 =4\fnl P_{xy}(k) \,,
\label{eq:pk_resp_phi}
\ee
where $(x,y)\in(\d,\phi)$. Combining \refeqs{pk_resp_d}{pk_resp_phi} with
\refeq{Phh}, we have
\ba
 \left.\frac{\partial\Phh(k)}{\partial\d_l}\right|_{\substack{{\rm linear}\\{\rm power}\\{\rm spectra}}}\:&=
 b_{10}^2\[\frac{47}{21}-\frac13\frac{d\ln\Pdd}{d\ln k}\]\Pdd(k)
 +b_{10}b_{01}\[\frac{47}{21}-\frac13\frac{d\ln\Pdd}{d\ln k}\]\Pdp(k) \,, \\
 \left.\frac{\partial\Phh(k)}{\partial\phi_l}\right|_{\substack{{\rm linear}\\{\rm power}\\{\rm spectra}}}\:&=
 4\fnl\Phh(k) \,.
\ea

The final way is through the reference halo number density. As the halo power
spectrum is computed referencing to the mean halo number density, in the presence
of long-wavelength fluctuations the mean halo number density measured by the
\emph{global} observer is a factor of $\(1+b_{10}\d_l+b_{01}\phi_l\)$ with respect
to that measured by the \emph{local} observer, and the halo power spectrum would
be rescaled by a factor of $\(1+b_{10}\d_l+b_{01}\phi_l\)^2$. At the leading
order, the halo power spectrum responds as
\be
 \left.\frac{\partial\Phh(k)}{\partial\d_l}\right|_{\substack{{\rm reference}\\{\rm density}}}=
 2b_{10}\Phh(k) \,, \quad
 \left.\frac{\partial\Phh(k)}{\partial\phi_l}\right|_{\substack{{\rm reference}\\{\rm density}}}=
 2b_{01}\Phh(k) \,.
\ee

Combining the three effects, the leading-order responses of halo power spectrum
to $\d_l$ and $\phi_l$ are
\ba
\label{eq:phh_resp1}
 \frac{\partial\Phh(k)}{\partial\d_l}\:&=
 2b_{10}b_{20}\Pdd(k)+b_{10}^2\[\frac{47}{21}-\frac13\frac{d\ln\Pdd}{d\ln k}\]\Pdd(k)+2b_{10}b_{11}\Pdp(k) \vs
 \:&+2b_{01}b_{20}\Pdp(k)+b_{10}b_{01}\[\frac{47}{21}-\frac13\frac{d\ln\Pdd}{d\ln k}\]\Pdp(k)+2b_{01}b_{11}\Ppp(k) \,, \\
 \frac{\partial\Phh(k)}{\partial\phi_l}\:&=
 2b_{10}b_{11}\Pdd(k)+2b_{01}b_{11}\Pdp(k)+2b_{10}b_{02}\Pdp(k)+2b_{01}b_{02}\Ppp(k) \vs
 \:&+4\fnl b_{10}^2\Pdd(k)+4\fnl b_{01}^2\Ppp(k)+8\fnl b_{10}b_{01}\Pdp(k) \,.
\label{eq:phh_resp2}
\ea
\refEqs{phh_resp1}{phh_resp2} are the main results of this section. We find
that they agree precisely with \refeqs{f_d}{f_phi}, demonstrating that the
halo squeezed-limit bispectrum can indeed be derived from considering how
halo power spectrum responds to the long-wavelength fluctuations.

\section{Discussion}
\label{sec:discu}
Using the SPT framework and the local Eulerian bias model, we show the
consistency between the halo squeezed-limit bispectrum and the responses
of the halo power spectrum to large-scale density and potential fluctuations.
Interestingly, for the perturbation theory one needs the second-order
computation to obtain the bispectrum, but for the response approach we
only have to consider how the \emph{linear} power spectrum responds to
large-scale fluctuations. Thus, it not only simplifies the computation
but also provides a novel way of understanding the physics of the squeezed-limit
bispectrum.

In this paper, we have demonstrated the responses using the perturbative
calculation, but the responses can readily be measured from separate
universe simulations to nonlinear scales. Specifically, in $\Lambda$CDM
cosmology the long-wavelength density fluctuation behaves as curvature
in the local universe, and one can perform $N$-body simulations in different
density environments to study how the matter power spectrum \cite{Li:2014sga,Wagner:2014aka}
and the halo mass function \cite{Lazeyras:2015lgp,Li:2015jsz,Baldauf:2015vio}
are affected. On the other hand, the local primordial non-Gaussianity
changes the amplitude of the local power spectrum, hence one can perform
simulations with different $\sigma_8$'s to study the effect on the halo
mass function \cite{Smith:2011ub,Biagetti:2016ywx}.

Combining with the separate universe simulations, the response approach is
powerful for exploring the observability as well as modeling the measurement
of the squeezed-limit bispectrum, especially in the era of blooming ongoing
and upcoming surveys. Here, we discuss two possible applications:
\begin{itemize}
{\item Construct a new model for the squeezed-limit bispectrum that works
better in the nonlinear regime. It has been shown in Ref.~\cite{Chiang:2014oga}
that the responses computed using nonlinear matter power spectrum models are
in better agreement with the matter squeezed-limit bispectrum measured from
simulations at $z\lesssim1$, compared to the second-order SPT. It can be
extended to halos, including the response of the mass function. Note, however,
that since galaxies are measured in redshift space with a preferred direction
exists, the response of the large-scale tidal field \cite{Dai:2015jaa,Ip:2016jji,Akitsu:2016leq}
has to be taken into account.}
{\item Predict the nonlinear squeezed-limit bispectrum formed by different
observables that cannot be computed by the perturbative approach, such as
the cross-correlation between the large-scale quasar overdensity and the
small-scale Lyman-$\alpha$ forest power spectrum \cite{Chiang:2017vsq}.
This is helpful for studying the constraining power on local primordial
non-Gaussianity using the squeezed-limit bispectrum of cross-correlation.}
\end{itemize}

Lastly, while we discuss the first-order response of the small-scale power
spectrum, one can generalize the calculation to the $m^{\rm th}$-order response
of the $n$-point function. For example, the $m^{\rm th}$-order response of the
small-scale power spectrum is equivalent to the $(m+2)$-point function with two
small- and $m$ large-scale modes \cite{Wagner:2015gva}, whereas the first-order
response of the small-scale $n$-point function is equivalent to the $(n+1)$-point
function with $n$ small- and one large-scale modes \cite{Adhikari:2016wpj}. The
response approach thus provides a novel and powerful technique to study the higher-order
statistics in the squeezed configurations for the large-scale structure.

\acknowledgements{
We thank Eiichiro Komatsu, Marilena LoVerde, and the referees for helpful
comments on the draft. CC is supported by grant NSF PHY-1620628.}

\appendix

\section{Halo squeezed-limit bispectrum from the second-order standard perturbation theory}
\label{app:bhhh}
In this appendix we derive explicitly the halo squeezed-limit bispectrum with
primordial non-Gaussianity from the second-order SPT. Using \refeq{d_h_3} and
the Wick's theorem, we obtain the leading-order bispectrum (up to $\delta^n\phi^m$
with $n+m=4$) in the squeezed limit such that $k_1\approx k_2\gg k_3$ as
\ba
 \:&\lim_{k_3\to0}\Bhhh(\vk_1,\vk_2,\vk_3)=\Bhhhsq(\vk_1,\vk_2,\vk_3) \vs
 =\:&b_{10}^2b_{20}[\Pdd(k_1)+\Pdd(k_2)]\Pdd(k_3)
 +2b_{10}^3[\Pdd(k_1)F_2(\vk_1,\vk_3)+\Pdd(k_2)F_2(\vk_2,\vk_3)]\Pdd(k_3) \vs
 +\:&b_{10}^2b_{11}[\Pdd(k_1)+\Pdd(k_2)]\Pdp(k_3)
 +b_{10}^2b_{11}[\Pdp(k_1)+\Pdp(k_2)]\Pdd(k_3) \vs
 +\:&b_{10}b_{01}b_{20}[\Pdp(k_1)+\Pdp(k_2)]\Pdd(k_3)
 +2b_{10}^2b_{01}[\Pdp(k_1)F_2(\vk_1,\vk_3)+\Pdp(k_2)F_2(\vk_2,\vk_3)]\Pdd(k_3) \vs
 +\:&b_{10}b_{01}b_{11}[\Ppp(k_1)+\Ppp(k_2)]\Pdd(k_3)
 +b_{10}b_{01}b_{11}[\Pdp(k_1)+\Pdp(k_2)]\Pdp(k_3) \vs
 +\:&b_{10}b_{01}b_{20}[\Pdd(k_1)+\Pdd(k_2)]\Pdp(k_3)
 +2b_{10}^2b_{01}[\Pdd(k_1)F_2(\vk_1,\vk_3)+\Pdd(k_2)F_2(\vk_2,\vk_3)]\Pdp(k_3) \vs
 +\:&b_{10}b_{01}b_{11}[\Pdd(k_1)+\Pdd(k_2)]\Ppp(k_3)
 +b_{10}b_{01}b_{11}[\Pdp(k_1)+\Pdp(k_2)]\Pdp(k_3) \vs
 +\:&b_{01}^2b_{20}[\Pdp(k_1)+\Pdp(k_2)]\Pdp(k_3)
 +2b_{10}b_{01}^2[\Pdp(k_1)F_2(\vk_1,\vk_3)+\Pdp(k_2)F_2(\vk_2,\vk_3)]\Pdp(k_3) \vs
 +\:&b_{01}^2b_{11}[\Pdp(k_1)+\Pdp(k_2)]\Ppp(k_3)
 +b_{01}^2b_{11}[\Ppp(k_1)+\Ppp(k_2)]\Pdp(k_3) \vs
 +\:&b_{10}^2b_{02}[\Pdp(k_1)+\Pdp(k_2)]\Pdp(k_3)
 +2\fnl b_{10}^2b_{01}[\Pdp(k_1)+\Pdp(k_2)]\Pdp(k_3) \vs
 +\:&2\fnl b_{10}^3[M(k_2)\Pdp(k_1)+M(k_1)\Pdp(k_2)]\Pdp(k_3)
 +b_{10}b_{01}b_{02}[\Ppp(k_1)+\Ppp(k_2)]\Pdp(k_3) \vs
 +\:&2\fnl b_{10}b_{01}^2[\Ppp(k_1)+\Ppp(k_2)]\Pdp(k_3)
 +2\fnl b_{10}^2b_{01}[M(k_2)\Ppp(k_1)+M(k_1)\Ppp(k_2)]\Pdp(k_3) \vs
 +\:&b_{10}b_{01}b_{02}[\Pdp(k_1)+\Pdp(k_2)]\Ppp(k_3)
 +2\fnl b_{10}b_{01}^2[\Pdp(k_1)+\Pdp(k_2)]\Ppp(k_3) \vs
 +\:&2\fnl b_{10}^2b_{01}[M(k_2)\Pdp(k_1)+M(k_1)\Pdp(k_2)]\Ppp(k_3)
 +b_{01}^2b_{02}[\Ppp(k_1)+\Ppp(k_2)]\Ppp(k_3) \vs
 +\:&2\fnl b_{01}^3[\Ppp(k_1)+\Ppp(k_2)]\Ppp(k_3)
 +2\fnl b_{10}b_{01}^2[M(k_2)\Ppp(k_1)+M(k_1)\Ppp(k_2)]\Ppp(k_3) \,,
\label{eq:Bhhh_1}
\ea
where we assume that $k_3$ is the long-wavelength mode and only contributes
to the linear-order perturbation, i.e. $\d_h(\vk_3)=b_{10}\d_l(\vk_3)+b_{01}\phi_l(\vk_3)$.
In the exact squeezed limit where $k_1=k_2=k$, the products of the linear
power spectra and the Poisson operator are
\be
 P_{\d\d}(k)=M(k_1)P_{\d\phi}(k_2)=M(k_2)P_{\d\phi}(k_1) \,, \quad
 P_{\d\phi}(k)=M(k_1)P_{\phi\phi}(k_2)=M(k_2)P_{\phi\phi}(k_1) \,.
\ee
Furthermore, taking the squeezed limit such that $k_3=k_L\to0$ as well as
angle-averaging the large-scale mode $\vk_3$, the combination of $F_2$ and
the small-scale power spectra at the leading order is given by (see e.g.
Ref.~\cite{Chiang:2014oga})
\be
 2[P_{xy}(k_1)F_2(\vk_1,\vk_3)+P_{xy}(k_2)F_2(\vk_2,\vk_3)]
 =\[\frac{47}{21}-\frac13\frac{d\ln\Pdd}{d\ln k}\]P_{xy}(k) \,,
\ee
where $(x,y)\in(\d,\phi)$. Combining the above equations, we can simplify
the halo squeezed-limit bispectrum with primordial non-Gaussianity as
\ba
 \:&\Bhhhsq(\vk_1,\vk_2,\vk_3) \vs
 =\:&\Bigg\lbrace 2b_{10}^2b_{20}\Pdd(k)
 +b_{10}^3\[\frac{47}{21}-\frac13\frac{d\ln\Pdd}{d\ln k}\]\Pdd(k)
 +2b_{10}^2b_{11}\Pdp(k) \vs
 \:&+2b_{10}b_{01}b_{20}\Pdp(k)
 +b_{10}^2b_{01}\[\frac{47}{21}-\frac13\frac{d\ln\Pdd}{d\ln k}\]\Pdp(k)
 +2b_{10}b_{01}b_{11}\Ppp(k)\Bigg\rbrace\Pdd(k_L) \vs
 +\:&\Bigg\lbrace 2b_{10}b_{01}b_{20}\Pdd(k)
 +b_{10}^2b_{01}\[\frac{47}{21}-\frac13\frac{d\ln\Pdd}{d\ln k}\]\Pdd(k)
 +2b_{10}b_{01}b_{11}\Pdp(k)
 +2b_{01}^2b_{20}\Pdp(k) \vs
 \:&+b_{10}b_{01}^2\[\frac{47}{21}-\frac13\frac{d\ln\Pdd}{d\ln k}\]\Pdp(k)
 +2b_{01}^2b_{11}\Ppp(k)
 +2b_{10}^2b_{11}\Pdd(k)
 +2b_{10}b_{01}b_{11}\Pdp(k)
 +2b_{10}^2b_{02}\Pdp(k) \vs
 \:&+2b_{10}b_{01}b_{02}\Ppp(k)
 +4\fnl b_{10}^3\Pdd(k)
 +4\fnl b_{10}b_{01}^2\Ppp(k)
 +8\fnl b_{10}^2b_{01}\Pdp(k)\Bigg\rbrace\Pdp(k_L) \vs
 +\:&\Bigg\lbrace 2b_{10}b_{01}b_{11}\Pdd(k)
 +2b_{01}^2b_{11}\Pdp(k)
 +2b_{10}b_{01}b_{02}\Pdp(k)
 +2b_{01}^2b_{02}\Ppp(k) \vs
 \:&+4\fnl b_{10}^2b_{01}\Pdd(k)
 +4\fnl b_{01}^3\Ppp(k)
 +8\fnl b_{10}b_{01}^2\Pdp(k)\Bigg\rbrace\Ppp(k_L) \,.
\label{eq:Bhhh_2}
\ea
In order to better compare with the linear halo power spectrum response,
it is useful to rewrite \refeq{Bhhh_2} as
\be
 \Bhhhsq(\vk_1,\vk_2,\vk_3)=b_{10}f_\d(k)\Pdd(k_L)+b_{01}f_\phi(k)\Ppp(k_L)+\[b_{01}f_\d(k)+b_{10}f_\phi(k)\]\Pdp(k_L) \,,
\ee
where
\ba
 f_\d(k)\:&=2b_{10}b_{20}\Pdd(k)
 +b_{10}^2\[\frac{47}{21}-\frac13\frac{d\ln\Pdd}{d\ln k}\]\Pdd(k)
 +2b_{10}b_{11}\Pdp(k) \vs
 \:&+2b_{01}b_{20}\Pdp(k)+b_{10}b_{01}\[\frac{47}{21}-\frac13\frac{d\ln\Pdd}{d\ln k}\]\Pdp(k)
 +2b_{01}b_{11}\Ppp(k) \,, \\
 f_\phi(k)\:&=2b_{10}b_{11}\Pdd(k)+2b_{01}b_{11}\Pdp(k)
 +2b_{10}b_{02}\Pdp(k)+2b_{01}b_{02}\Ppp(k) \vs
 \:&+4\fnl b_{10}^2\Pdd(k)+4\fnl b_{01}^2\Ppp(k)+8\fnl b_{10}b_{01}\Pdp(k) \,.
\ea

\bibliography{draft}

\end{document}